\documentclass[12pt]{article}
\usepackage{graphicx}
\usepackage[T1]{fontenc}

\begin{document}

\title{Cosmological applications of a geometrical interpretation of "c".}

\author{J.-M. Vigoureux, P. Vigoureux, B. Vigoureux\\
Institut UTINAM, UMR CNRS 6213,\\Université de
Franche-Comté, 25030 Besançon Cedex,France.\\
{\em jean-marie.vigoureux@univ-fcomte.fr}}

\maketitle \fontsize{12}{24}\selectfont

\begin{abstract}
We make the hypothesis that the velocity of light and the expansion of the universe are two aspects of one single concept connecting space and time in the expanding universe. We show that solving Friedman's equations with that interpretation (keeping $c$ = constant)
could explain number of unnatural features of the standard cosmology.
We thus examine in that light the flatness and the quintessence problems, the problem of the observed uniformity in
term of temperature and density of the cosmological background
radiation and the small-scale inhomogeneity problem. We finally show that using this interpretation of $c$ leads to reconsider the Hubble diagram of distance moduli and redshifts as obtained from recent observations of type Ia supernovae without having to need an accelerating universe.
\end{abstract}

keywords : {cosmology, cosmic background radiation,
cosmology:observation, cosmological parameters, flatness, speed of
light, accelerating universe, small-scale inhomogeneity, large-scale smoothness}

\section{Introduction}

The constant $c$ was first introduced as the speed of light.
However, with the development of physics, it came to be understood
as playing a more fundamental role, its significance being not
directly that of a usual velocity (even though its dimensions are)
and one might thus think of $c$ as being a fundamental constant of
the universe. Moreover, the advent of Einsteinian relativity, the
fact that $c$ does appear in phenomena where there is neither light
nor any motion (for example in the fundamental equation $E$ = $m$
$c^2$) and its double-interpretation in terms of "velocity of light"
and of "velocity of gravitation" forces everybody to associate $c$
with the theoretical description of space-time itself rather than
that of some of its specific contents; it invites us to connect $c$
to the geometry of the universe. In this paper, we propose to
connect $c$ to the expansion of the universe and we show some
consequences of this interpretation in cosmology. We thus examine the flatness problem and the large-scale smoothness problem in the light of that interpretation of $c$. In the third part, we also show that these results are compatible with the observation of small-scale inhomogeneity leading to the structures seen by astronomers. We finally show that this model doesn't need to consider an accelerating universe to interpret the Hubble diagram of distance moduli and redshifts as obtained from recent observations of type Ia supernovae.

\section{A geometrical interpretation of $c$}

The universal constant $c$ sets up {\em a universal relation
between space and time}. This relation is expressed in the
relativistic invariant $ds^2$ which gives on the light cone $c =
dx/dt$. On the other hand, the expansion of the universe {\em provides another universal relation between space and time} which also
has the physical dimension of a velocity.

These two results cannot express a fortuitous coincidence: on the
contrary, as in our previous works \cite{vig01,vig03}, we consider as a logical necessity that there are not {\em
two different} universal relations between space and time
having the same physical dimension of a velocity. We thus suggest
that the velocity of light $c$ and the expansion of the universe are two aspects of one single concept connecting space and time in the expanding universe. Taking
this into account, we can write in the simplest possible form (of
course, all the following results also holds when taking $c=1$)

\begin{equation}\label{eq:c}
c=K \frac {dR(t)}{dt}
\end{equation}

\noindent where $K$ is a positive constant. It must
be noted that here "$c$" and consequently $dR/dt$
is constant (no acceleration or deceleration). That choice could
seem to lead to a model which fails to describe the real universe.
In fact, \emph{in the usual cosmology}, a constant velocity of
expansion needs either a totally empty universe (which is obviously
not the case) or a cosmological constant exactly canceling matter
density and pressure \emph{at all times} which is not fulfilled in
usual theories. However, it will be shown in part IV that using
eq.(\ref{eq:c}) in the second Friedmann's equation can lead to
other solutions.

In \cite{vig01,vig03} we have suggested that c must
be the recessional velocity of our antipode (that is to say the
variation with respect to time of half the spatial circumference of
the universe at cosmological time t) so that $c=\pi dR(t)/dt $. This choice of $c$ makes it to play the role of $\pi$ in euclidian geometry and underlines its geometrical nature. It also makes the recessionnal velocity of the observer's antipode to be $c$. We choose this value in what follows but another numerical choices
for $K$ would lead to similar results.

Eq.(\ref{eq:c}) leads to a new physical interpretation of the
Einstein's constant $c$ which thus appears to be related to the
variation of R(t) with time. It shows that $c$ can be defined from
the knowledge of the geometry of space-time only, that is from its
size and its age. It thus really takes the statute of a true
geometrical fundamental magnitude of the universe.

\section{Some cosmological consequences of that interpretation of $c$}
\subsection{the flatness problem}

Our first aim is to show how Eq.(\ref{eq:c}) can solve the flatness problem. The question has been asked to know how $\Omega$ (which is the ratio of the mass/energy density of the universe) has been so highly fine-tuned in the past to give an approximately flat universe. We show that with this hypothesis on $c$, the universe \textsl{must appear to be flat whatever it may be} (spherical or not) so that it is not surprinsing that it is found to be flat.

Noting that, using $c$=constant, the two 
Friedmann equations remain valid. The first one is

\begin{equation}\label{eq:friedmann}
\left( \frac{\dot{R}(t)}{R(t)}\right) ^2 =\frac{8\pi G \rho }{3}
-\frac{k\,c^2}{R(t)^2}+ \frac{\Lambda\,c^2}{3},
\end{equation}

\noindent where $k$ is the curvature parameter. R(t) is related to the scale
factor a(t) by the relation $R(t)$ = $R_0$ $a(t)$ where the
subscript 0 refers to a quantity evaluated at the present time.

In the case of a {\em flat} universe ($k=0$) and {\em when using the
standard model}, eq.(\ref{eq:friedmann}) reduces to
\begin{equation}\label{eq:friedmannplat}
\left( \frac{\dot{R}(t)}{R(t)}\right) ^2 =\frac{8\pi G \rho }{3}+
\frac{\Lambda\,c^2}{3},
\end{equation}

In the case of a {\em spherical} universe $(k = 1)$ and {\em when
using the above interpretation of $c$} (that is to say when using
eq.(\ref{eq:c})), eq.(\ref{eq:friedmann}) becomes (the writing of the last term is not modified in order to be clearer in what follows) 
\begin{equation}\label{eq:friedmannplat1}
\left( \frac{\dot{R}(t)}{R(t)}\right) ^2 =\frac{8\pi G \rho }{3} -
\pi^2 \left( \frac{\dot{R}(t)}{R(t)}\right) ^2 +
\frac{\Lambda\,c^2}{3},
\end{equation}
and consequently becomes
\begin{equation}\label{eq:flatness}
\left( \frac{\dot{R}(t)}{R(t)}\right) ^2 = \frac{8\pi G}3 \frac{\rho
}{\pi^2 +1}+ \frac{\Lambda\,c^2}{3 (\pi^2 +1)}.
\end{equation}

So, eq.(\ref{eq:flatness}) obtained {\em with $k=1$} when using the present
interpretation of "$c$" is quite similar to eq.(\ref{eq:friedmannplat})
obtained {\em with $k=0$} in the standard model. This shows that
when using eq.(\ref{eq:c}) a spherical universe would
\textit{experimentally} appear to be flat \textit{but with a smaller mass} than expected in standard cosmology since the mass density is $\rho/(\pi^2 +1)$ in eq.(\ref{eq:flatness}) whereas it is $\rho$ in eq.(\ref{eq:friedmannplat}). This explains why it appears to be natural to {\em observe} a
flat universe even if the probability for such an universe is nearly
null because of instability.

\subsection{the large-scale smoothness problem}
Let us now consider the large-scale smoothness problem in the light of eq.(\ref{eq:c}). As everybody knows, the cosmic background radiation (CBR) is amazingly uniform across regions in opposite directions. When looking, for example, at two distant regions on opposite sides of our universe, the observed temperature only fluctuates to about one part per 100. Eq.(\ref{eq:c}) radically changes the causal structure of the universe and explains simply this observation. In fact, it shows that it is the {\em same} small part of the past universe that is observed in all the possible directions of space so that it is no more surprising to find such a thermal and a density uniformity. 

Using both the Robertson-Walker metric and eq.(\ref{eq:c}) with $K = \pi$ (but, note again that another numerical value of K can be taken without changing essential results), the
horizon is obtained by calculating

\begin{equation}\label{eq:horizon}
\psi_h = \int_ {t_h} ^{t_0} c \frac {dt}{R(t)})= \pi \int_ {R(t_h)}
^{R(t_0)} \frac {dR(t)}{R(t)})= \pi \ln(\frac {R(t_0)}{R(t_h)})
\end{equation}

In the present model, there exists values of $R(t_h)$ giving
$\psi_h > \pi$ so that there is no object horizon. This can be seen
by noting that when using eq.(\ref{eq:c}), all the observed lines of
sight of an observer cut themselves in the past so that {\em there
exists an event which can be seen in any directions around us}. This
can be computed by using $ds^2 = 0$ along a light ray. It gives the space-time of a given present observer shown in fig.(\ref{fig1}). Solving

\begin{equation}\label{eq:ds2}
- c^2 dt^2 + R(t)^2 (\frac{dr^2}{1 - k r^2}) \equiv - c^2 dt^2 +
R(t)^2 d\psi^2 = 0
\end{equation}

\noindent with eq.(\ref{eq:c}) gives

\begin{equation}\label{eq:spiral}
R(t)=R(t_0)\exp(-\frac{\psi} {\pi } )\
\end{equation}

As a consequence of eq.(\ref{eq:spiral}), eq.(\ref{eq:c}) makes the
space-time of a given present observer to be closed on itself at an early time: for $\psi=\pi$,
there is an "isotropic event", that is denoted A in fig.~\ref{fig1},
which can be seen in any direction around us and which can be
identified to the source of the cosmic background radiation. Since
it is thus the {\em same} event A of the early universe that is seen
in any directions (the cosmic microwave background radiation arriving at
the earth from all directions in the sky does come from the {\em same} small
part of the early universe), it is not surprising to observe a very
high uniformity in terms of temperature and of density of the CBR.

Note that eq.(\ref{eq:spiral}) also shows that other "isotropical
events" (we mean events which could be observed in any direction
around us) could exist. They are defined by $\psi= n \pi$ (where
$n$ is an integer). They could correspond to blackbody radiations at
temperature other than $2.74$ $K$ or to particles other than photons
such as cosmic particles. However these other isotropic events can
also be non visible because of the opacity of the universe at early
times.

\subsection{the small-scale inhomogeneity}
A related comment concerns the problem of the small-scale inhomogeneity needed to explain the formation of all the observed structures of the universe. Eq.(\ref{eq:c}) shows that the homogeneity of the CBR is compatible with such inhomogenities and consequently with the existence of structures of all scales. In fact, as a consequence of the previous part, the CBR constitutes only a very small part of the past universe and consequently all its other parts (which cannot be seen at the same past age), can be quite different.    

\begin{figure}[h]
\centering
\includegraphics[width=10cm]{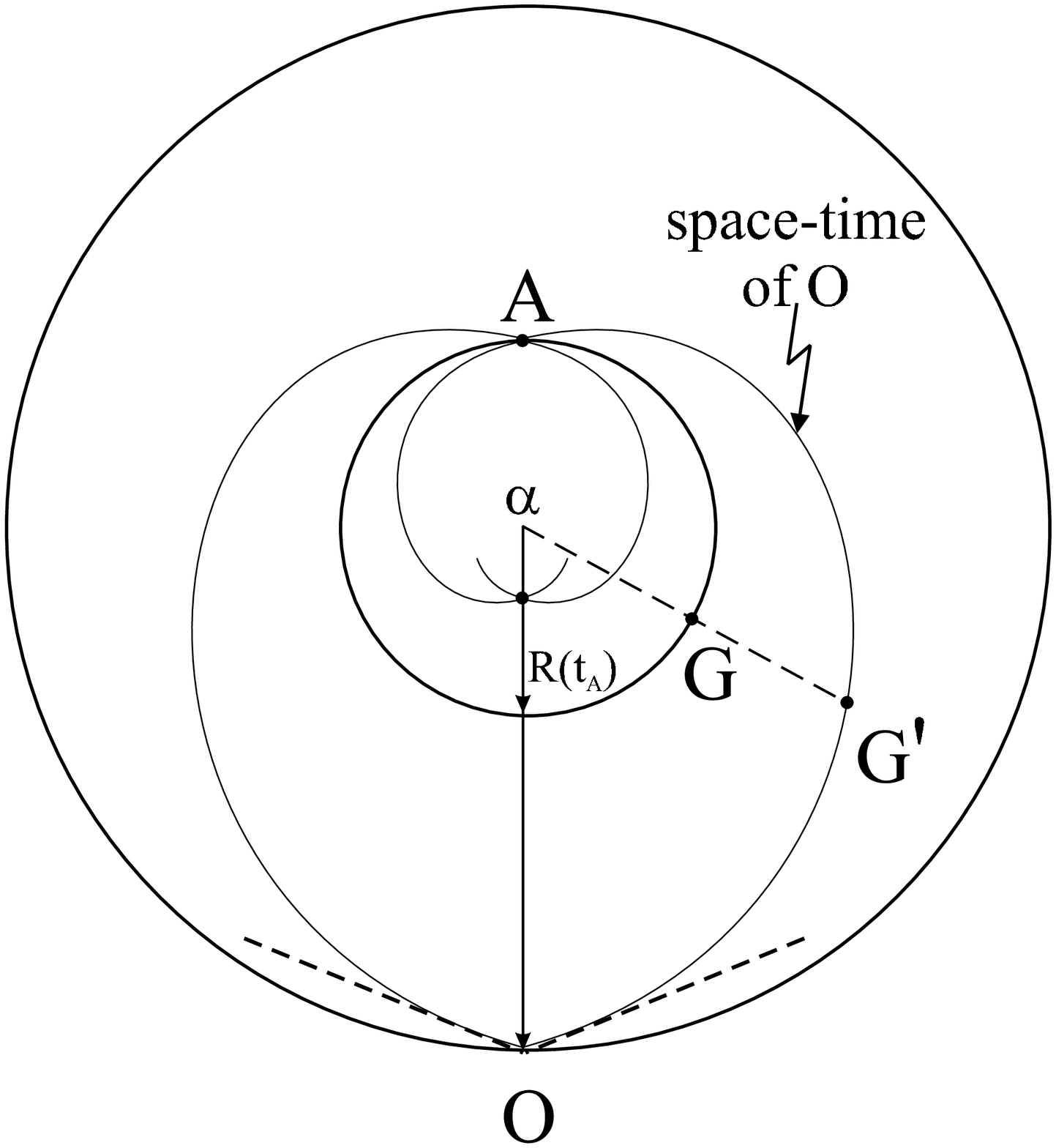} \caption{The space-time of an observer O when using 
eq.(\ref{eq:c}). $\alpha$ corresponds to the big-bang.
Cosmological time corresponds to the radius line $\alpha-O$. Space
(for each given cosmological time $t$) corresponds to circles. The
space-time of an observer O is constructed by drawing the light cone
from O by going back into time and by taking into account the
connection of $c$ with the expansion of the universe. It can be
seen that, as one goes back in time, the light cone area spreads out
to greater distance and then decreases toward the
big-bang. Near O, it corresponds to the usual light cone (dashed
lines). Globally, it is closed on events A, A'... which can be
observed {\em in any direction around us}. Among them, A can be identified
to the source of the cosmic background radiation. Since it is the
same very small volume A of the early universe which is observed in
any directions, we have not to be surprised to measure a very high
isotropy and homogeneity of the cosmic background radiation (CBR). The
universe, at the time $t_{\rm A}$ that the CBR was emitted, is
illustrated by the circle the radius of which is $R(t_{\rm A})$. A
thus corresponds to only a very small element of the universe at
time $t_{\rm A}$. The observed isotropy is consequently not
contradictory with the existence at this same time $t_{\rm A}$ of
inhomogeneities (such as G). However, such inhomogeneities cannot be
seen at time $t_{\rm A}$ (since only A belongs to the space-time of the observer O at
that time) but only at G' at later time $t_{\rm G'}$. } \label{fig1}
\end{figure}

Eq.(\ref{eq:spiral}) with $\psi=\pi$, shows that,
at cosmical time $t_{\rm A}$, $R(t_{\rm A}) \neq0$. Because of this, the source
A of the cosmological background radiation in fig.~\ref{fig1}
constitutes only a very small volume element of the universe at that
time. The existence of A is consequently consistent with that of
spatial inhomogeneities (anywhere else in the universe at the same
time $t_{\rm A}$ that the radiation was emitted). Such
inhomogeneities (for example G on the figure), may physically
coexist with A at time $t_{\rm A}$ without being observable at that
time (among all the volume elements of the universe at time $t_{\rm
A}$ only the one noted A on fig.~\ref{fig1} is in the present
space-time of the observer $O$ and can consequently be seen). They can be the original
seeds giving birth, at later times, to galaxies and other structures
which are now observed at $t_{\rm G'}>t_{\rm A}$ (G' on fig.~\ref{fig1}).

Let us note that this example also shows that it could be possible
to see, just behind a galaxy $G'$ (or exactly in the opposite
direction), but at an earlier time, the cosmical object which has
given birth to that galaxy. Moreover, let us also note that, because
of the spiral form of the light rays which are wound round the
big-bang, such objects could seem to be older than the value
corresponding to the present cosmic time.

\subsection{the Hubble diagram of distance moduli and redshifts}
\begin{figure}
\centering
\includegraphics[width=12cm]{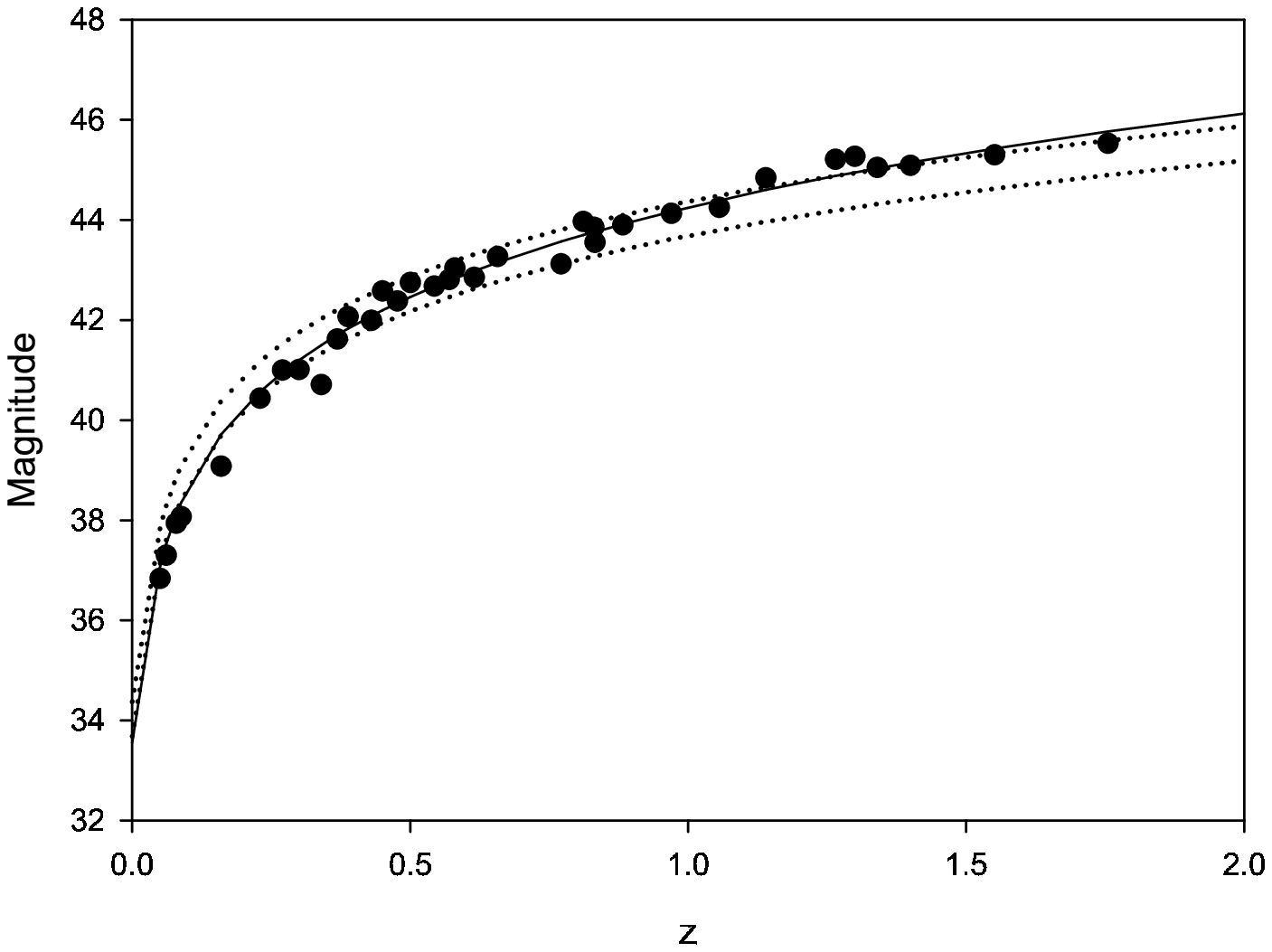}
\caption{Hubble diagram (distance modulus vs redshift). The
data points are taken from Table 5 of ref \cite{rie04}. The two dashed lines show the results obtained with eq.(\ref{eq:hubblestandard}). One is obtained by fitting the small values of z ($z < 1$). The other is obtained when fitting the high values of z ($z > 1$). They show that it is not possible to fit experimental data {\em both for $z < 1$ and for $z > 1$ in the standard model}. 
The full line shows that observations can be fitted both for $z < 1$ and
for $z > 1$ by using eq.(\ref{eq:hubble}) (and consequently by using
eq.(\ref{eq:c})). The full line which represents predictions
of the present model has been drawn by using $H_0 = 58 km.s^{-1}.Mpc^{-1}$.} \label{fig2}
 \end{figure}

Another problem is the one of the accelerating universe. A very distant supernova with a redshift $z = 1.755$ was recently observed by the Hubble Space Telescope. In the standard cosmology, that observation seems to show that the universe is accelerating. In fact, usual models fail to fit that new data point at redshift 1.755. It can be shown that eq.(\ref{eq:c}) leads to another expression for the distance-moduli which can fit this point with a good precision (see fig.~\ref{fig2})

As is well known, the standard expression for
the corrected distance-moduli with respect to z can be written (ref.\cite{Weinberg72})

\begin{equation}\label{eq:hubblestandard}
\mu = 25 + 5 \log cz -5\log H_0 + 1.086 (1 -q_0)z + ...
\end{equation}

Following pioneering works related in ref.\cite{Nor89}, recent
observations of type Ia supernovae with the Hubble Spatial Telescope
(refs.\cite{wang03,tonry03,rie04,sch04}) has provided a robust extension of the
Hubble diagram to $1 <z <1.8$. These new results have shown that
observations {\em cannot be fitted by using
eq.(\ref{eq:hubblestandard}) both for $z < 1$ and for $z > 1$}. To
solve this problem, some authors have concluded for the necessity of
ruling out the traditional $(\Omega_M,\Omega_\Lambda)=(1,0)$
universe. The universe would have consequently been decelerating
before the current epoch of cosmic acceleration.

When using eq.(\ref{eq:c}) it can be shown that, within the present model, the expression for
the corrected distance-moduli is given no more by
eq.(\ref{eq:hubblestandard}), but by
\begin{equation}\label{eq:hubble}
\mu = 25 + 5\log(\frac {c}{H(t_0)}) + 5 \log(1+z) + 5\log \ln (1+z)
\end{equation}

\noindent where $c$ is in km.s$^{-1}$ and $H$ in km.s$^{-1}$.Mpc$^{-1}$.
As shown on fig.~\ref{fig2} that result permits to fit the
experimental values in the whole range $z < 1$ and for $z > 1$
without any other considerations. Thus, the use of eq.(\ref{eq:c}) succeeds in fitting all the data without having to consider acceleration.

\section{On the general aspect of the geometry}

Let us now consider the second Friedmann equation:

\begin{equation}\label{eq:friedmann2}
\left( \frac{\ddot{R}(t)}{R(t)}\right) =-\frac{4 \pi G}{3} (\rho + 3
\frac{p}{c^2})+ \frac{\Lambda\,c^2}{3}.
\end{equation}
By using eq.(\ref{eq:c}) (recalling that $c$=constant and neglecting $p$) that equation leads to
\begin{equation} \label{eq:densities}
\Lambda c^2 \approx 4 \pi G \rho.
\end{equation}

This result can explain why the density of cosmological constant
($\Lambda$ $c^2$)/($8$ $\pi$ $G$) is so near the matter density (it
shows why $\rho_{V}$ is not only small but also, as current Type Ia
supernova observations seem to indicate, of the same order of
magnitude as the present mass density of the universe). To understand this result note that eq.(\ref{eq:densities}) leads to

\begin{equation}\label{eq:lambda2}
\Lambda = \frac{4 \pi G \rho}{c^2}= \frac{G M}{R
 c^2}\frac{3 }{R^2}
\end{equation}

\noindent Introducing eq.(\ref{eq:lambda2}) with eq.(\ref{eq:c}) into eq.(\ref{eq:flatness}) also gives

\begin{equation}\label{eq:lambda1}
\Lambda = \frac{\pi^2 +1}{\pi^2} \frac{1}{R^2}.
\end{equation}

Eq.(\ref{eq:lambda2}) and eq.(\ref{eq:lambda1}) can be solved as

\begin{equation}\label{eq:mach}
\frac{G M}{R c^2}= Constant,
\end{equation}
and
\begin{equation}\label{eq:lambdaR-2}
\Lambda \approx \frac{1}{R^2}
\end{equation}

It is interesting to note that
eq.(\ref{eq:mach}) is an expression of the Mach's principle (as for
example discussed in refs.\cite{vig03,costa99,costa00,Sciama53},
whereas the eq.(\ref{eq:lambdaR-2}) has been shown to be in
conformity with quantum cosmology and with existing observations
by Chen Wei and Wu Yong-Shi (ref.\cite{chen90}). So, using
eq.(\ref{eq:c}) with $c$ constant can lead to a coherent model (more general solutions could be obtained by using a possible variation of $c$ with respect to time)

\section{Conclusions}

 We propose that "$c$" is intimately connected to the expansion of
the universe. This interpretation gives to $c$ a geometrical meaning and
makes of it a true {\em universal} quantity of the universe which can be defined from its size and its age without any other considerations.

It can solve the problem of flatness: the flat model is
unstable; even slight deviations from flatness grow very quickly,
leading inevitably, to either a catastrophic big crunch or
emptiness. This conflicts with the fact that observations show that
the universe is so near flatness. The above interpretation of $c$ solves that problem by showing that the universe {\em must appear to
be flat} (and with a smaller mass than that expected in the standard cosmology) {\em whatever may be its geometrical form}. It can also give a
new light at the horizon problem by explaining the homogeneity and
the isotropy of the observed universe at early times and it makes
possible the coexistence with the isotropic background radiation of
inhomogeneities which can be the seeds of galaxies or of other cosmical
objects. It may also permit the observation behind galaxies (or in
the opposite direction) of the cosmical object which has given birth
to it.

This interpretation of $c$ also leads to a good fitting (without having to consider an accelerating expansion) of observations in Hubble diagram of
distance modulus with respect to redshifts as obtained from recent
observations of type Ia supernovae.

In concluding, let us underline that usual values obtained from observations in the field of the usual standard cosmology cannot be too quickly transposed to judge the present model. All observations are in fact to be reinterpreted within the present model.

Let us also add that connecting the constant $c$ to the expansion of
the universe and more precisely, to the geometry of the universe,
may also clarify some intriguing properties of the so-called "speed
of light"' which appears in phenomena in which
neither light nor any motion is present, which also appears in so
different fields of physics that are electromagnetism and
gravitation and which has both the interpretation of "velocity of
light" and of "velocity of gravitation".

\bibliographystyle{unsrt}

\clearpage

 \clearpage

 \clearpage

\clearpage

\end{document}